\documentclass[a4paper]{article}

\input{pkg.def}

\input{biblio.def}

\input{math.def}

\title{Do irrotational water waves remain irrotational in the limit of a vanishing viscosity?}
\author{Alan Riquier\,\orcidlink{0009-0005-2179-924X}\thanks{Corresponding author. Email address: \href{mailto:alan.riquier@ens.psl.eu}{\texttt{alan.riquier@ens.psl.eu}}} }
\author{Emmanuel Dormy\,\orcidlink{0000-0002-9683-6173}}
\affil{Département de Mathématiques et Applications,
    CNRS UMR 8553,\\ 
    École Normale Supérieure - PSL, 45 rue d'Ulm, 75005 Paris, France}
\date{}

\begin{document}

\maketitle

\begin{abstract}
    Theoretical results on water waves almost always start by assuming irrotationality of the flow in order to simplify the  formulation. In this work, we investigate the well-foundedness of this hypothesis via numerical simulations of the free-surface Navier-Stokes equations. 
    
    We show that, in the presence of a non-flat bathymetry, either angular or smooth, a gravity wave of finite amplitude can shed vortex pairs from the bottom boundary layer into the bulk of the flow. As these eddies approach the free surface they modify the shape of the wave. {It is found that this perturbation does not vanish as the Reynolds number is increased. The vanishing viscosity limit of water waves is therefore singular when no-slip boundary conditions are enforced on the bottom}.
\end{abstract}

\section*{Introduction}

The theoretical description of water waves is one of the oldest and most challenging topics in fluid dynamics as well as in wave theory \cite[ch. 51]{feynman_1965}. This remains an area of intense research, with key open questions still unresolved -- such as wave breaking and wave turbulence. The highly non-linear nature of the phenomenon, combined with the fluid's very low viscosity, severely limits the use of the full governing equations in both analytical and numerical studies. As a result, simplified models are commonly used in practice. Such models include the shallow water equations \cite{SaintVenant1871}, the Green-Naghdi system \cite{Green_Naghdi_1976} or the K--dV \cite{KdV1895} equation and its generalisation to non-flat topographies \cite{Johnson_1973}. They all correspond to asymptotic regimes of the full water waves system \cite{Zakharov1968, CraigSulem1993,johnson,lannes2013}, which is itself derived from the Euler equations assuming that the flow is irrotational for all times. Although this assumption is widely accepted, it is rarely justified. Here, we examine its validity.

Though small, the effect of viscosity is to yield boundary layers at the bottom and near the free-surface, which were observed in numerical works \cite{Raval2009, Deike2015, Riquier_Dormy_2024} as well as experimentally \cite{grue_2017}. In the absence of spilling or breaking, the free surface contributes little to the vorticity generation (the associated boundary layer having a spatial extent scaling as $\nu^{1/2}$, with $\nu$ the kinematic viscosity, and a finite amplitude vorticity correction as the viscosity vanishes). The bottom boundary layer however remains relevant in the limit of vanishing viscosity (it also extends over a size $\nu^{1/2}$ but involves a vorticity correction of amplitude $\nu^{-1/2}$). As the viscosity is decreased, convergence to the irrotational solution can only hold if this vorticity sheet becomes infinitely localised on the boundary \cite{kato_1984, kelliher_2007}. In that case, the shape of the free-surface continuously tends to that of the Euler flow. 

In 1953, Longuet-Higgins \cite{LonguetHiggins1953} raised the question of the stability of the water bed vorticity sheet:
\begin{quotation}
    \textit{\say{In an oscillating motion, this vorticity will be of alternating sign; and the question then presents itself: will any of the vorticity spread into the interior of the fluid, or will it remain in the neighbourhood of the boundary?}} 
\end{quotation}
Here we address this question and demonstrate, through a numerical study of the incompressible free-surface Navier-Stokes equations, that a wave passing over an obstacle can result in vorticity being shed from the boundary layer into the bulk. Owing to the oscillatory nature of the flow, we observe the formation of counter-rotating vortex pairs (or modons) within the fluid, whose effects on the free surface do not decrease as the Reynolds number is increased. The vanishing viscosity limit is thus singular for this problem. Numerical experiments are conducted using both angular and smooth topographies, and it is argued that a Reynolds number-dependent curvature threshold exists, above which separation of the vorticity sheet happens.

\section{Mathematical formulation and numerical approximation}

We numerically solve the full non-linear Navier–Stokes equations with a free surface \cite[see][]{Riquier_Dormy_2024}, considering a non-flat bathymetry where the no-slip boundary condition is applied. The fluid domain $\Omega(t)$ is two-dimensional, periodic in the $x$-direction with period $L$, bounded below by a rigid bottom $\Gamma_b$ and above by a moving interface $\Gamma_i(t)$. We introduce $\vect{\gamma}_b(s)$ a parametrisation of $\Gamma_b$ and $\vect{\gamma}_i(t,s)$ a time-dependent parametrisation of $\Gamma_i(t)$. To facilitate the smoothing of the topography in Section \ref{sec:smoothness}, we further assume that $\vect{\gamma}_b(s)$ corresponds to the arc-length parametrisation. A schematic representation of this configuration is shown in Fig. \ref{fig:def}.

\begin{figure}[ht]
    \centering
    \begin{tikzpicture}[scale=2.1]
        \fill[color=cyan!20]
                (0,0) -- (0,1.5) 
                -- plot [domain=0:2*pi,samples=200] (\x,{1+0.2*cos(\x r + pi r)+0.08*sin(2*\x r)}) 
                -- (2*pi,0) -- (4*pi/3,0)
                -- (4*pi/3,0.5) -- (2*pi/3,0.5) 
                -- (2*pi/3,0) -- cycle;
        \fill[color=gray!50] (2*pi/3,0) -- (4*pi/3,0)
                -- (4*pi/3,0.5) -- (2*pi/3,0.5) 
                -- cycle ;
        \draw[thick] (2*pi,0)
                --  (4*pi/3,0)
                --  (4*pi/3,0.5) -- (2*pi/3,0.5) 
                --  (2*pi/3,0) -- (0,0);
        \draw[->] (4*pi/3,0) -- (2*pi+0.25,0) node[below right]{$x$} ;
        \draw (0,0) -- (2*pi/3,0);
        \draw[dashed] (2*pi/3,0) -- (4*pi/3,0);
        \draw[->] (0,0) -- (0,1.5) node[above left]{$y$} ;
        \draw (2*pi,0) -- (2*pi,1.2) ;
        \draw[thick] plot[domain=0:2*pi,samples=200](\x,{1+0.2*cos(\x r + pi r)+0.08*sin(2*\x r)});
        \draw[dashed] (0,1) -- (2*pi,1);
        \draw[|-|] (-0.1,0) -- (-0.1,1) node[midway,left]{$h_0$};
        \draw (1,0.5) node{$\Omega(t)$};
        \draw (4.2*pi/3,0.25) node{$\Gamma_b$};
        \draw (3.9,1.4) node{$\Gamma_{i}(t)$};
        \draw[|-|] (0,-0.1) -- (2*pi,-0.1) node[midway,below]{$L$};
        \draw[|-|] (2*pi/3,0.6) -- (4*pi/3,0.6) node[midway, above]{$\ell_b$};
        \draw[|-|] (1.9*pi/3,0) -- (1.9*pi/3,0.5) node[midway, left]{$h_b$};
        
        \draw (-0.05,0.4) -- (0.05,0.5);
        \draw (-0.05,0.425) -- (0.05,0.525);
        \draw (2*pi-0.05,0.4) -- (2*pi+0.05,0.5);
        \draw (2*pi-0.05,0.425) -- (2*pi+0.05,0.525);
    \end{tikzpicture}
    \caption{Geometry of the domain $\Omega(t)$ in dimensional units. The bathymetry can be chosen arbitrarily. It is pictured here as a rectangular step of height $h_b$ and length $\ell_b$ as it shall be used in sec. \ref{sec:rect_step}.}
    \label{fig:def}
\end{figure}

The domain $\Omega(t)$ is filled with a fluid (usually water) whose kinematic viscosity is denoted by $\nu$. We neglect the effects of an upper fluid (usually air) on the dynamics of the flow (in effect setting its density to zero). We also disregard the effects of surface tension, even though they could easily be included in this formulation. The characteristic length of the flow is chosen as the distance $h_0$ between the average height of the interface and the minimum height of the bottom (see Fig. \ref{fig:def}). The chosen typical velocity $U$ corresponds to the velocity of an inviscid gravity wave in shallow water, \textit{i.e.} $U = \sqrt{gh_0}$. The non-dimensional governing equations then take the form
\begin{subequations}\label{eq:Navier-Stokes}
    \begin{equation}
        \frac{\partial\vect{u}}{\partial t} + (\vect{u}\cdot\n)\vect{u} = -\n p + \frac{1}{Re}\Delta\vect{u} - \hat{\vect{y}}\, ,
    \end{equation}
    \vspace*{-1em}
    \begin{equation}
        \n\cdot\vect{u} = 0\, ,
    \end{equation}
\end{subequations}
where the Reynolds number defined as
\begin{equation}\label{eq:Re}
    Re = \frac{h_0\sqrt{gh_0}}{\nu} \, ,
\end{equation}
and  $\hat{\vect{y}}$ is the unit vector along the $y$-axis (Fig. \ref{fig:def}). From now on, all quantities appearing in this work should be understood as non-dimensional.

The no-slip condition 
\begin{equation}\label{eq:no-slip}
    \vect{u} = 0
\end{equation}
is  imposed on the bottom boundary $\Gamma_b$. As we shall see, the formation of a boundary layer near $\Gamma_b$ can have a significant effect on the overall flow and on the interface.

On the free-surface $\Gamma_i(t)$ the stress-free condition,
\begin{equation}
    \label{eq:stress-free}
    p \, \hat{\vect{n}} - \frac{1}{Re} \Big[ \n\vect{u} + \big(\n\vect{u}\big)^t \Big] = 0,
\end{equation}
is enforced. Each fluid particle on the interface is advected as a Lagrangian tracer. This is achieved through the kinematic condition
\begin{equation}
    \frac{\partial \vect{\gamma}_i}{\partial t} (t,s) = \vect{u}\big(t,\vect{\gamma}_i(t,s)\big).
\end{equation}

Our results will be  compared to those obtained solving the irrotational Euler equation. To that end, we rely on the code described in \cite{DormyLacave2024}. The Euler equation being only first-order in space, one cannot use the same set of boundary conditions to close the inviscid free-surface system. The stress-free condition is sufficiently well-behaved to converge gently to the dynamic condition $p = 0$ in the limit of a vanishing viscosity (see \cite{Masmoudi_Rousset_2017} for a mathematical proof of this statement for a wave of small amplitude and the numerical work \cite{Riquier_Dormy_2024} for evidences that it still holds up to the moment the free surface self-intersects). On the bed, however, things are different. Indeed, the free-surface Euler system is closed using the non-penetrability condition $\vect{u}\cdot\hat{\vect{n}}_b = 0$ only and one cannot enforce $\vect{u}\cdot\hat{\vect{\tau}}_b = 0$ too (with $\hat{\vect{n}}_b$ and $\hat{\vect{\tau}}_b$ the unit vectors respectively perpendicular and tangent to the bottom boundary). What happens to this boundary condition in the limit of high Reynolds number is the subject of the present work. Indeed, boundary layer separation may happen and prevent the convergence to the free-surface Euler system.

In both cases, an irrotational initial condition is chosen. The initial interface $\vect{\gamma}_i(t=0)$ is chosen as a simple cosine wave of finite amplitude $A$. The initial velocity $\vect{u}(t=0)$ is built from an initial potential $\phi_0$ computed by solving Laplace's equation. The normal derivative of $\phi_0$ on the initial free surface corresponds to the first order solution of the water waves equations (see \textit{e.g.} \cite{johnson}, chap. 2),
\begin{equation}
   \partial_n \phi_0 \big|_{\Gamma_i(0)} = \vect{u}_0 \cdot \hat{\vect{n}}\quad\text{with}\quad \vect{u}_0 = A \sqrt{gk\tanh(kh_0)}\cdot \begin{bmatrix}
        \big(\tanh(kh_0)\big)^{-1}\cos(kx) \\ \sin(kx)
    \end{bmatrix},
\end{equation}
where $g = h_0 = 1$ in non-dimensional units and $\hat{\vect{n}}$ is the unit vector perpendicular to $\Gamma_i(0)$ pointing outside of $\Omega(0)$. 

The solution to this evolutionary problem is computed numerically using the FreeFEM finite-element code on unstructured meshes \cite{FreeFem}. A crank-Nicolson time-stepping scheme is used for all terms but the non-linearity, which is treated in a mixed explicit-implicit manner in order to prevent the use of a non-linear solver. The time step is recalculated at each time iteration using the smallest element of the unstructured mesh, in order to satisfy the CFL condition. In order to advect the interface points with the velocity of the fluid, we employ the arbitrary Lagrangian-Eulerian method \cite{ALE_HIRT1974}. The entire mesh is advected at each time step with a velocity $\vect{w}$, obtained after numerically solving the following problem
\begin{equation}
    \left\{ \begin{array}{rcll}
        \Delta\vect{w} &=& 0 & \text{in the fluid } \Omega(t) \, ,\\
        \vect{w} &=& \vect{u} & \text{on the interface } \Gamma_i(t) \, ,\\
        \vect{w} &=& 0 & \text{on the bottom } \Gamma_b \, .
    \end{array} \right. 
\end{equation}
This is a Lagrangian advection scheme for the interface points while preventing spurious triangle elongation near the bottom boundary layer. An Adaptative Mesh Refinment (AMR) algorithm is used periodically to fully resolve the vortical structures numerically. Numerical convergence of this method is discussed in appendix \ref{sec:appendix}.

\section{Separation from a rectangular step}
\label{sec:rect_step}

\begin{figure}[ht]
    \centering
    \includegraphics[width=\textwidth, trim=0cm 0cm 1cm 0cm, clip]{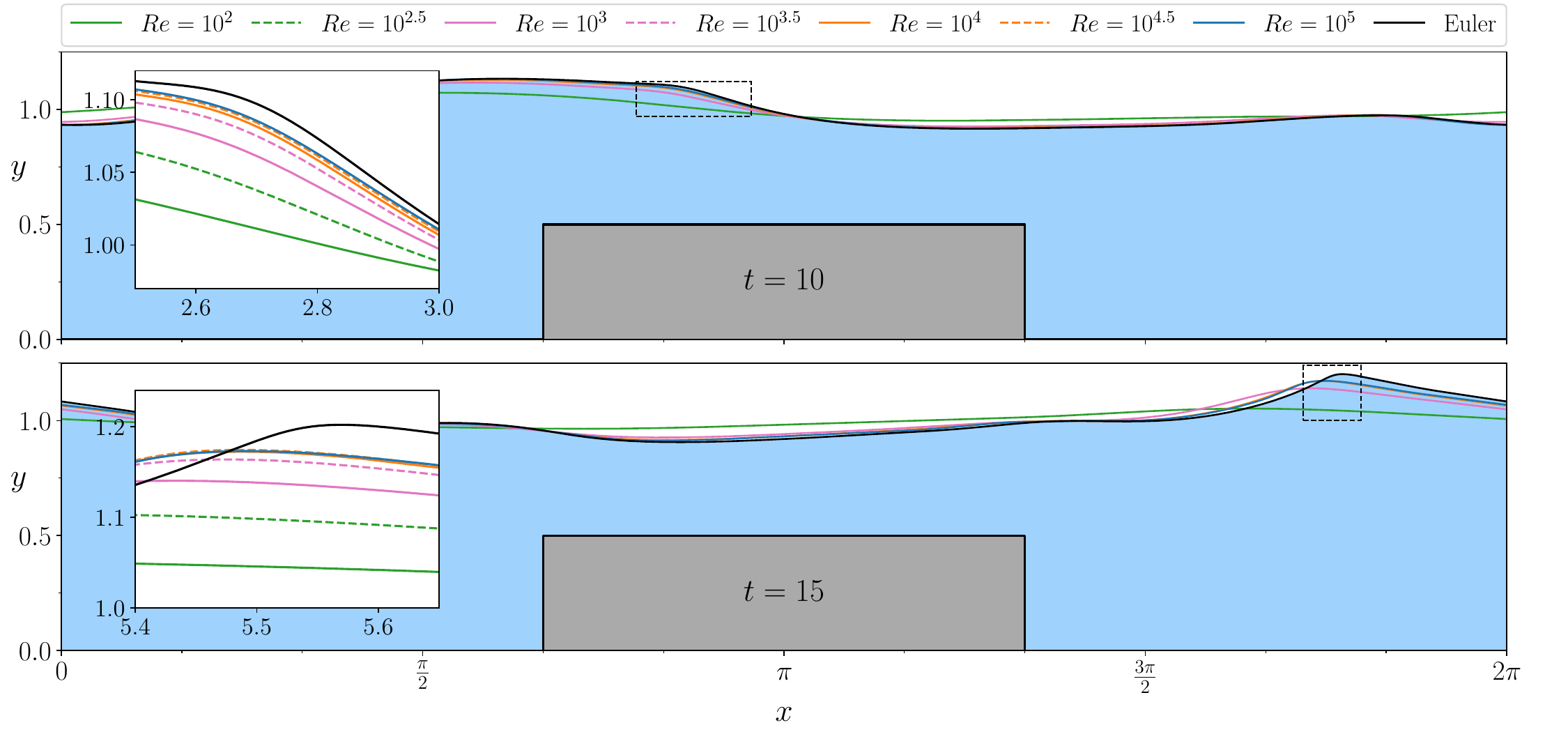}
    \caption{Snapshots of the interfaces for different values of $Re$ and the free-surface irrotational Euler solution, at times $t = 10$ and $t = 15$. In all these simulations, the initial condition has been computed setting $L = 2\pi$, $A = 0.1$, $k=1$, $h_b = 0.2$ and $\ell_b = L/3$. A significant difference remains in the limit of small viscosity. The Euler simulation has been obtained using the code of Dormy and Lacave \cite{DormyLacave2024}. The numerical parameters used to obtain this result are given in table \ref{tab:numerical_params}.}
    \label{fig:compare_interfaces}
\end{figure}

We investigate a classical setup  originally introduced by Lamb \cite[§176]{lamb_1932}, and also often used experimentally \cite[\textit{e.g.}][]{grue_1992} or numerically \textit{e.g.} to study soliton fission \cite[see][for instance]{viotti_dutykh_dias_2014}, in which the bathymetry is flat except for a rectangular step, here of {(non-dimensional)} length $\ell_b = {2\pi}/{3}$ and height $h_b = 0.5$ (see fig. \ref{fig:def}). The wave's initial amplitude is chosen as $A = 0.1$, large enough to ensure a non-negligible flow velocity in the vicinity of the bed, but small enough to prevent wave breaking. Its initial wavenumber is $k=1$ so that the domain's (non-dimensional) length $L = 2\pi$ corresponds to one wavelength. Numerical simulations have been carried out with different values of Reynolds' number, using the numerical parameters shown in table \ref{tab:numerical_params}. The resulting interface evolution for different values of $Re$ and the irrotational solution are shown in Fig. \ref{fig:compare_interfaces}. The solutions display an interesting and unexpected behaviour: a finite gap between the viscous and inviscid solutions persists, even as the viscosity is reduced.

In order to investigate whether the Navier-Stokes solution might converge to the Euler solution as the Reynolds number is increased, we need to introduce a quantitative measurement of the distance between two solutions. Since the interface remains a graph, we can compute the $L^2$ distance between two interfaces. Let $\eta_E(t,x)$ the function whose graph represents the irrotational free surface at time $t$. Similarly, for each value of $Re$, introduce the function $\eta_{Re}(t,x)$ representing the free surface. To quantify the gap between the viscous and inviscid solutions, we introduce the two  distances,
\begin{equation*}
    \big(d_E(t;Re)\big)^2 = \int_{\mathbb{T}} \big( \eta_E(t,x) - \eta_{Re}(t,x) \big)\,\mathrm{d}x\quad\text{and}\quad \big(d_5(t;Re)\big)^2 = \int_{\mathbb{T}} \big( \eta_{10^5}(t,x) - \eta_{Re}(t,x) \big)\,\mathrm{d}x,
\end{equation*}
for the values of $Re$ ranging from $10^2$ to $10^5$ visible in Fig. \ref{fig:compare_interfaces}, and with $\mathbb{T} = [0,2\pi)$ the one-dimensional torus.

\begin{figure}[ht]
    \centering
    \includegraphics[width=\textwidth, trim=0cm 0cm 0cm 0cm, clip]{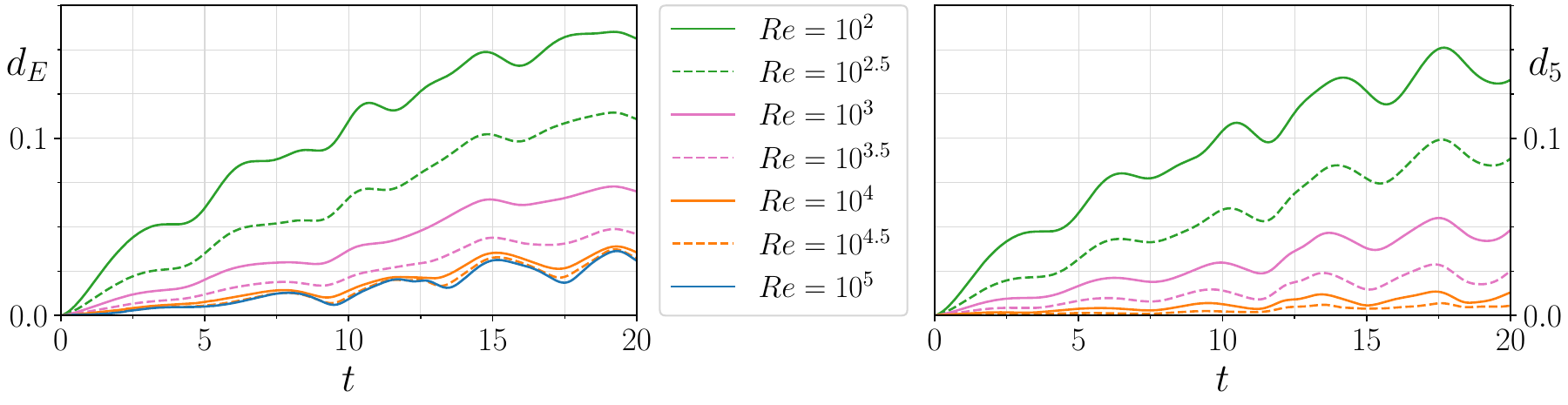}
    \caption{On the left panel, time evolution of the $L^2$ distance $d_E$ between the free surface obtained from the Navier-Stokes solution and the one corresponding to the inviscid irrotational Euler solution for different values of $Re$. The distance $d_5\, ,$ with the simulation whose Reynolds number is $Re = 10^5  ,$ is presented on the right.}
    \label{fig:L2_distance}
\end{figure}

The time evolution of both $d_E$ and $d_5$, computed numerically from the simulations described previously, is shown in Fig. \ref{fig:L2_distance}. On the one hand, fig. \ref{fig:L2_distance}(\textit{left}), we see that the distance $d_E$ between the viscous and inviscid free surfaces converges, as $Re$ is increased, to a non-vanishing value which slowly increases with time, on average. The distance $d_5$ (fig. \ref{fig:L2_distance}, \textit{right}) to the highest Reynolds number, on the other hand, decreases with $Re$. Hence the viscous solution converges to a limit interface which differs from the irrotational inviscid one. This is the case, despite the fact our initial condition was irrotational. As we shall see below, this arises from the emergence of vortices within the fluid domain, which persist in the vanishing-viscosity limit and give rise to finite perturbations of the interface.

\begin{figure}[ht]
    \centering
    \includegraphics[width=\textwidth, trim=0cm 0cm 0cm 0cm, clip]{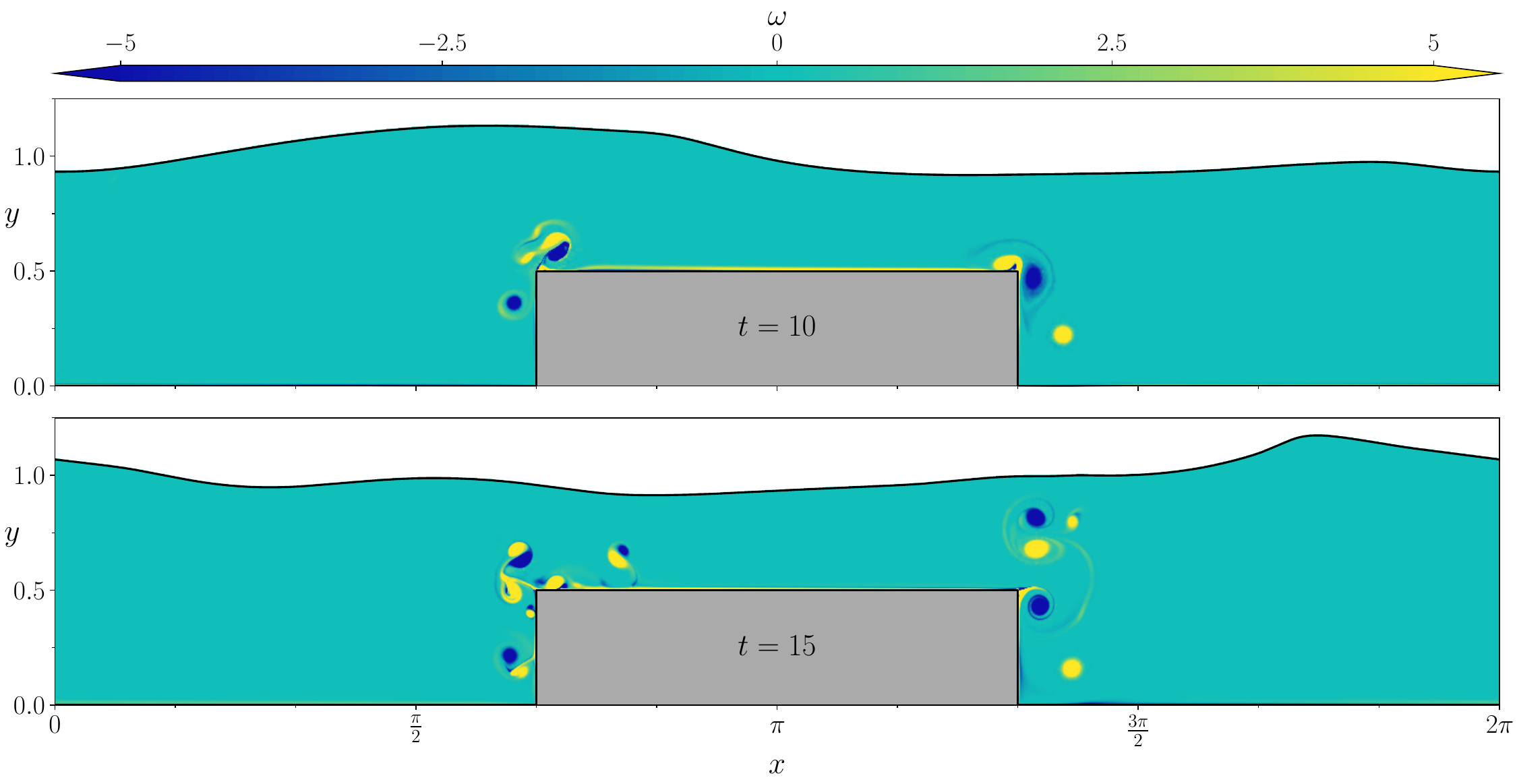}
    \caption{Evolution of the vorticity at $Re = 10^5$ (movie available as supplementary material). The color code is centred on $[-5,5]$ to highlight eddies in the flow.}
    \label{fig:vortices_Re5}
\end{figure}

The origin of this singular limit can be traced back to the vorticity $\omega = \partial_x u_y - \partial_y u_x\, ,$ as presented in Fig.~\ref{fig:vortices_Re5} for $Re = 10^5$. The velocity induced by the passing wave triggers an instability in the bottom boundary layer. In this way, vorticity is continuously shed into the bulk, with an intensity that depends on the background (irrotational) velocity of the flow. Such vorticity filaments have also been observed in the numerical and experimental studies of Lin and Huang \cite{lin_huang_2010}. They rapidly roll up on themselves due to the Kelvin–Helmholtz instability, giving rise to the vortex-like structures visible in Fig. \ref{fig:vortices_Re5}. Accordingly, we shall refer to these vortical regions as \say{vortices} hereafter. The oscillatory nature of the flow, away from the boundary layer at the bed, leads to the formation of counter-rotating vortex pairs, or modons, through the mechanism described above.

These vortex pairs  occasionally travel toward the free surface and at other times toward the bottom, where they can extract vorticity (see {the left part of} Fig. \ref{fig:vortices_Re5} at time $t=15$ as well as \cite{nguyen_etal_2011}). 

\begin{figure}[ht!]
    \centering
    \includegraphics[width=\textwidth, trim=0cm 0cm 0cm 0cm, clip]{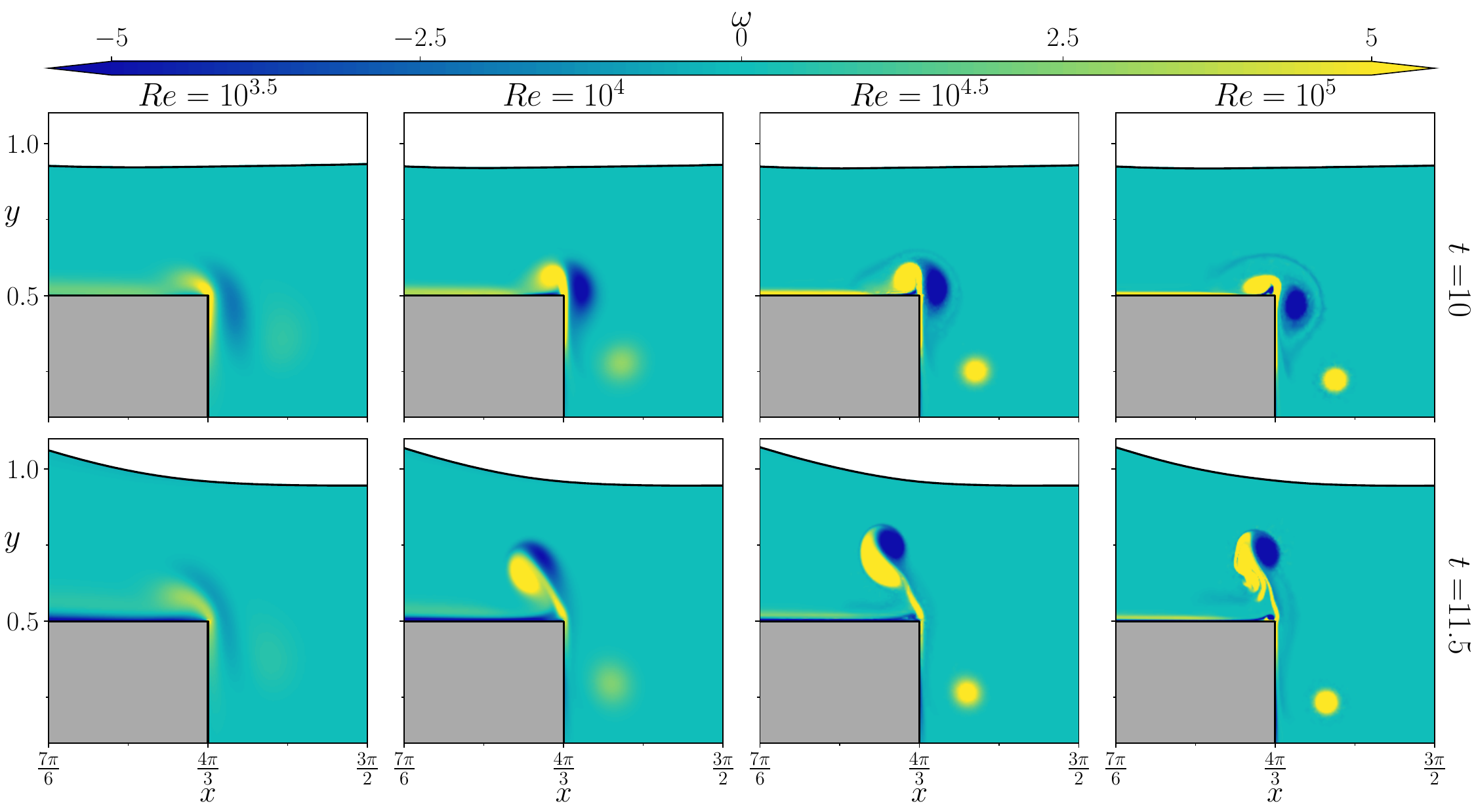}
    \caption{Vorticity field at $t = 10$ and $t=11.5$ for Reynolds numbers $Re$ ranging from $Re = 10^{3.5}$ to $Re = 10^5$.}
    \label{fig:vortices_comp}
\end{figure}

A comparison between the solutions for various values of $Re$ at time $t = 10$ is shown in Fig. \ref{fig:vortices_comp}. Modons being rapidly propagating vortex pairs, their effect extends far from the bottom boundary and can rapidly reach the free surface. Their contribution does not vanish with the viscosity, which prevents the free-surface to converge to the irrotational Euler solution as $Re$ tends to infinity. Instead the solution appears to converge toward an irrotational flow except for localised vorticity in the bulk \cite{lagerstrom_1975}.

\begin{figure}[ht!]
    \centering
    \includegraphics[width=\textwidth]{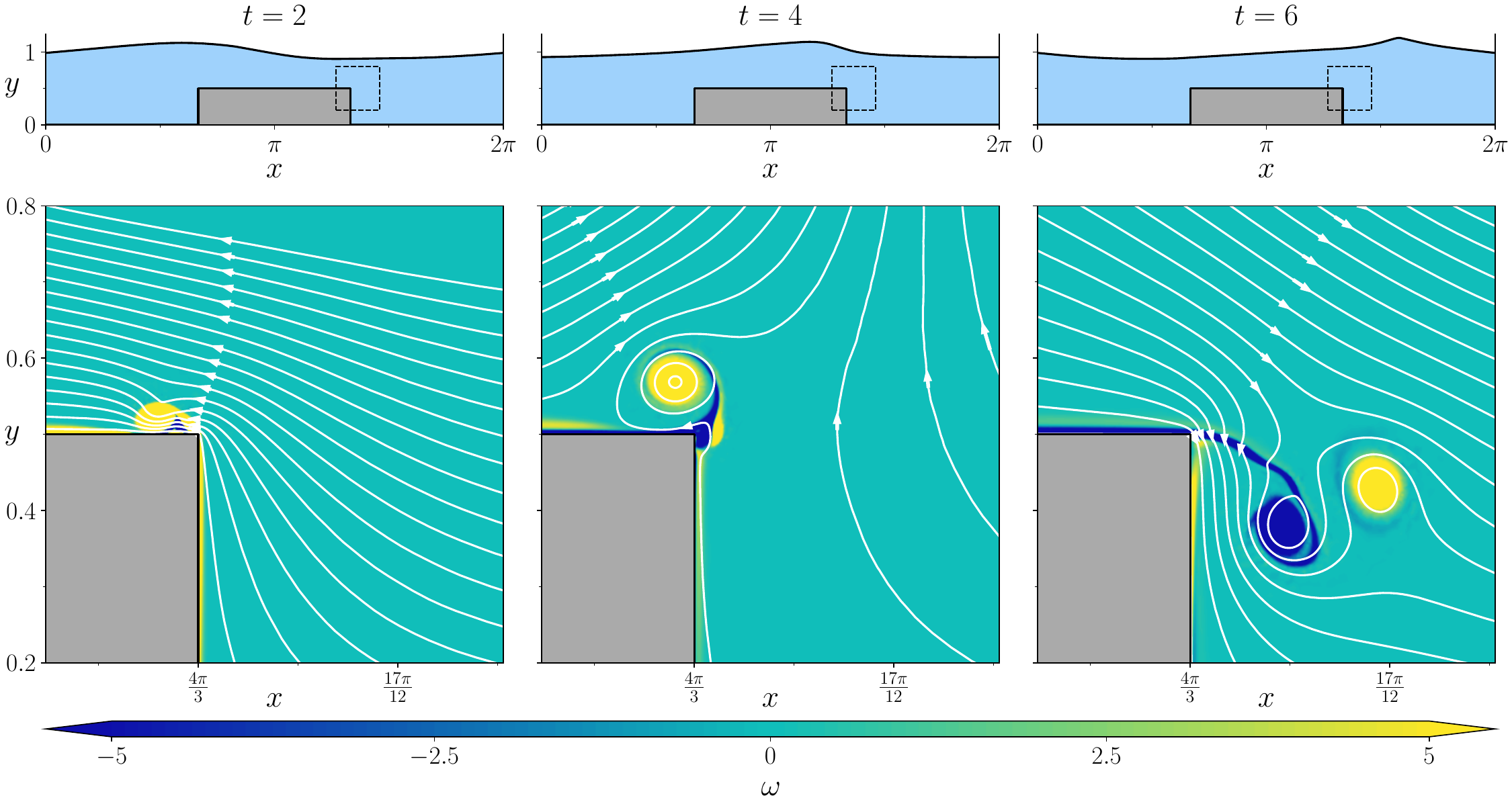}
    \caption{Streamlines (white lines with arrows) and vorticity fields $\omega$ near the top right corner of the obstacle at $t = 2$(left), $4$(middle) and $t = 6$(right) for $Re = 10^5$.}
    \label{fig:stream}
\end{figure}

\begin{figure}[ht!]
    \centering
    \includegraphics[width=\textwidth]{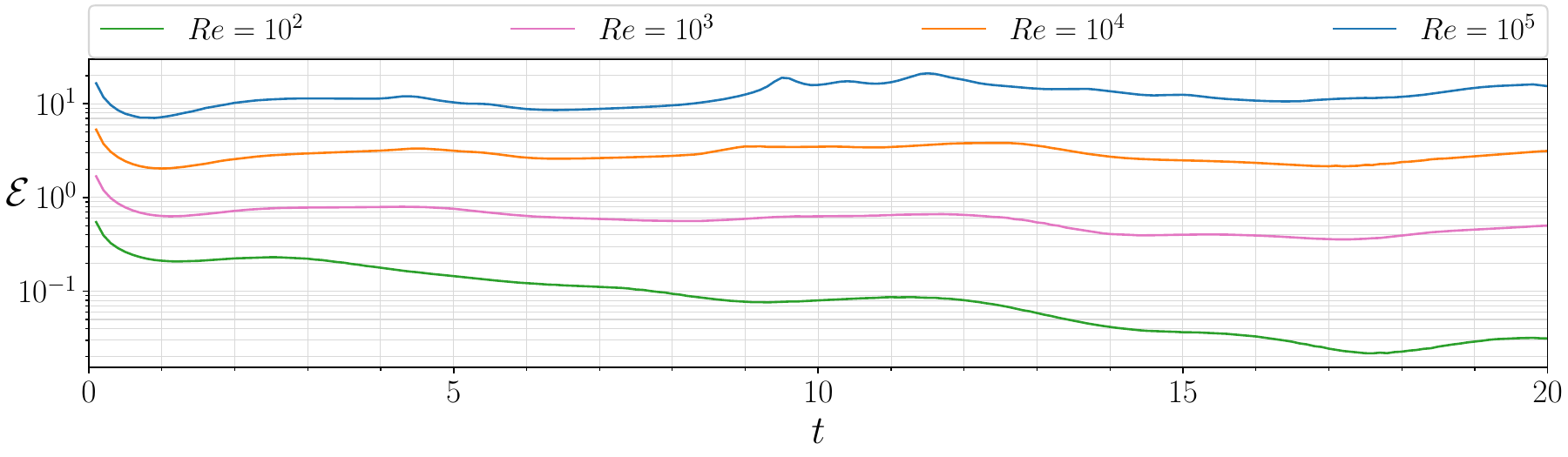}
    \caption{Evolution of the enstrophy $\mathcal{E}(t;Re)$, defined in eq. \eqref{eq:enstrophy}, for different values of the Reynolds number $Re$.}
    \label{fig:ensrophy}
\end{figure}

The drag force on submerged topography associated with vorticity generation is well established both in observations and in experiments \cite{Hasselmann68,Lowe07,lin_huang_2010}. The vortex-generating mechanism can be understood by considering streamlines near the trailing corner as the first pair of vortices is generated, see Fig.~\ref{fig:stream}. As expected, the streamlines wrap around the crests and the troughs of the wave. Locally, in the vicinity of the obstacle the underlying irrotational flow gets through three different stages. First, near $t = 2$ it corresponds to an up-going flow near a corner, whose velocity potential is well-known \cite[see \textit{e.g.}][sec. 6.5]{Batchelor_1967}. At a later time (about $t=4$), as the inflection point of the interface passes above the salient edge, the direction of the stream changes. The streamlines no longer wrap around the topography. Instead the horizontal left-coming flow meets the vertical up-going one at the corner in a fashion similar to that of the rear tip of an aircraft's wing. These streamlines effectively guide the pair of vortices out of the boundary layer. At later times ($t = 6$ in fig. \ref{fig:stream}), the underlying irrotational flow wraps once again around the corner, but now in the opposite direction.

The singular limit highlighted above follows from the  formation of vortex pairs, or modons, which induce a finite perturbation of the wave profile, even in the limit of vanishing viscosity. The formation of modons follows from a combination of the direction of the mainstream and vorticity stripping from the boundary layer. Throughout a crest, the direction of the flow outside the boundary layer corresponds to the wave propagation direction and the vorticity generated in the boundary layer has a negative sign (fig. \ref{fig:stream} right). As a trough passes above the edge, the direction of the stream is opposite to the wave propagation so that the vorticity has a positive sign (fig. \ref{fig:stream} left). Thus negative (resp. positive) vorticity is stripped from the boundary layer during the passing of crests (resp. troughs). 

We can quantify the vorticity generated by the bed's boundary layer at the global level by computing the enstrophy $\mathcal{E}$, defined as
\begin{equation}\label{eq:enstrophy}
    \mathcal{E}(t;Re) = \int_{\Omega(t)} \omega^2(t,\vect{x})\, \mathrm{d}\vect{x},
\end{equation}
where the dependency of both $\omega(t,\vect{x})$ and $\Omega(t)$ in the Reynolds number $Re$ is implicit. The integral \eqref{eq:enstrophy} is evaluated numerically to yield the enstrophy evolution shown in Fig. \ref{fig:ensrophy} for different values of $Re$. Its value remains stable through the course of a simulation. It scales linearly with the Reynolds number $Re$, thus preventing the viscous dissipation to vanish in the limit $Re\to +\infty$. For large enough Reynolds numbers the vorticity stripped from the boundary layer balances the loss due to viscosity.

\section{Separation from a smooth topography and finite curvature}
\label{sec:smoothness}

We now want to investigate whether vorticity stripping persists around edges of finite curvature. To that end, we consider smoothed out versions of the topography by using the method of mollifiers \cite{friedrichs_1944_mollifiers}.  The function $\vect{\gamma}_b(s) \, ,$ i.e. the arclength parametrisation of the rectangular step considered in sec. \ref{sec:rect_step}, is non-differentiable at the edges. In order to make it smooth and differentiable, we introduce the following bump function,
\begin{equation}
    \phi(x) = \begin{cases}
        C\exp\Big( - \frac{1}{1 - x^2} \Big) & \text{for } -1 < x < 1 \, ,\\
        0 & \text{otherwise,}
    \end{cases}
\end{equation}
where $C$ is a normalisation factor ensuring that the integral of $\phi$ from $x = -1$ to $x = 1$ is set to unity. This function is infinitely differentiable. Let us introduce a parameter $\varepsilon > 0$ which will account for the \textit{curvature radius} of the obstacle. The mollifier $\phi_\varepsilon$ is then defined as
\begin{equation}
    \phi_\varepsilon(x) = \frac{1}{\varepsilon} \cdot \phi\left( \frac{x}{\varepsilon} \right),
\end{equation}
and can be used to produce a smooth bathymetry using a convolution 
\begin{equation}
\vect{\gamma}_\varepsilon(s) = \int_{-\varepsilon}^{\varepsilon} \phi_\varepsilon(t)\,\vect{\gamma}_b(s-t)\,{\rm d}t \, ,
\end{equation}
which converges to $\vect{\gamma}_b(s)$ as $\varepsilon\to 0$.

\begin{figure}[ht]
    \centering
    \includegraphics[width=\textwidth, trim=0cm 0cm 0cm 0cm, clip]{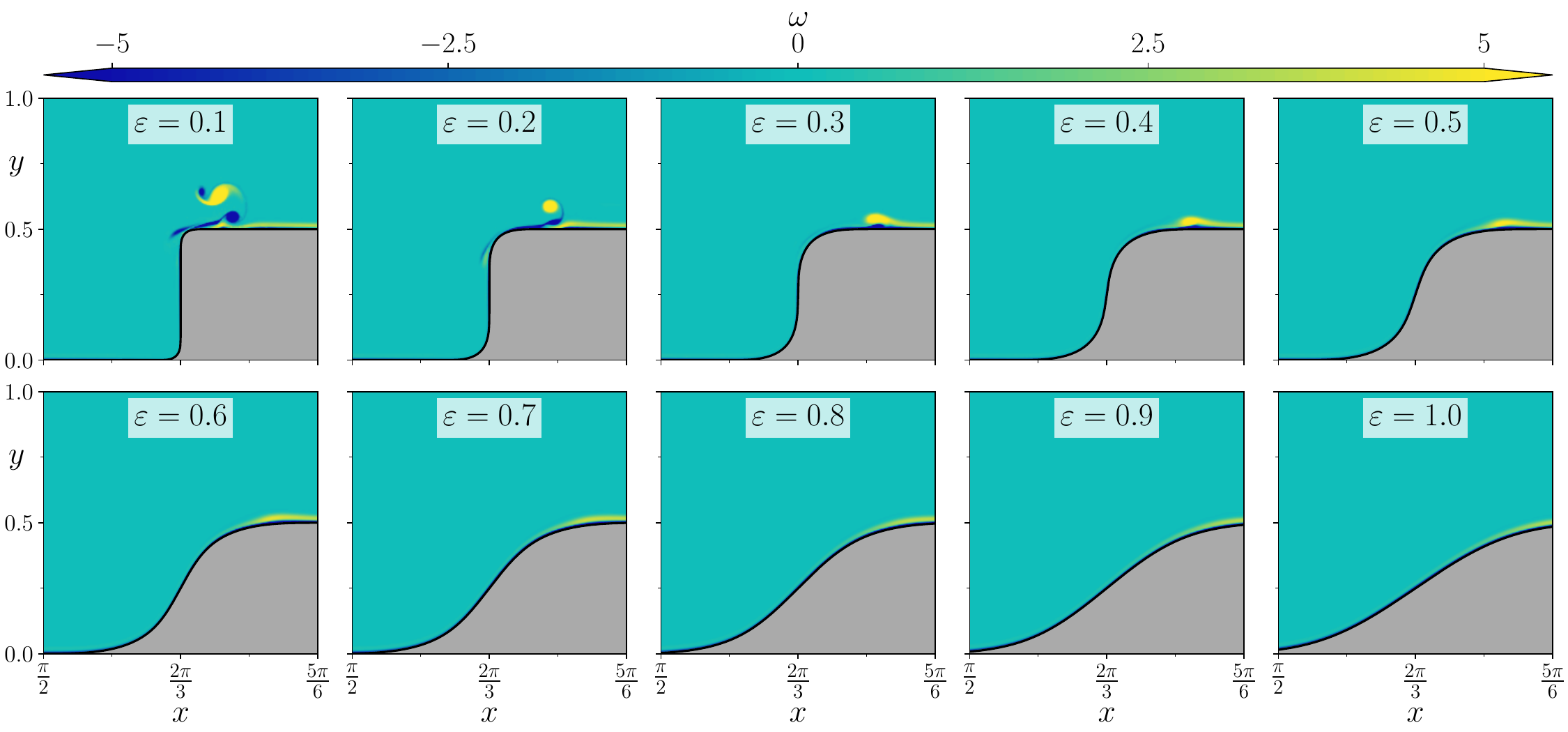}
    \caption{Emission of vortices {at $Re = 10^5$} and $t=10$ in the vicinity of the left side of the obstacle as its smoothness $\varepsilon$ varies. In each numerical simulation presented here, the initial condition has been computed using the values $L = 2\pi$, $A = 0.1$, $k = 1$, $h_b = 0.5$ and $\ell_b = L/3$ (before mollification), as in sec. \ref{sec:rect_step}. The numerical parameters used to obtain this result are written down in table \ref{tab:numerical_params}.}
    \label{fig:obstacle_curvature}
\end{figure}

For values of $\varepsilon$ ranging from $0.1$ to $1$, the associated vorticity generation is presented in Fig. \ref{fig:obstacle_curvature}. At $Re = 10^5$, for small $\varepsilon$, vortex pairs are still being emitted into the fluid. As $\varepsilon $ is increased, at fixed $Re$ the boundary layer is stabilized.

The presence of these vortices is reminiscent of the von Kármán \cite{vonKarman1911} vortex street, in which an adverse pressure gradient in the boundary layer leads to the emission of vortices past a cylinder when the Reynolds number is increased. We are, in fact, considering a rather similar situation except that {the separation} is not here associated with a steady stream, but instead results from an oscillation of the background flow in time.

A boundary layer Reynolds number $Re_b$ can be defined from the length of the step $\ell_b$ and the velocity of the background potential flow in the mainstream $U_b$. For a first-order Stokes wave over an almost-flat topography, we have $U_b \approx A \cdot \sqrt{{2gk}/{\sinh(2kh_0)}}.$ As the viscosity $\nu$ tends to zero, $Re_b \to + \infty$ and separation must happen when a critical value of $Re_b$ is achieved \cite{lagerstrom_1975,sychev_etal_1998}. Denoting as $R_c$ the minimum curvature radius of the topography, a classical scaling argument for inviscid irrotational flows yields that ${U_b} \sim O\{ R_c^{-1} \}$ as $R_c\to 0$ and therefore $Re_b \to +\infty$ in this limit. Hence, for each finite Reynolds number $Re$, a critical curvature radius $R_c$ will yield separation. The aspect ratio of the submerged obstacle is not expected to be essential to modon emission, unlike the flow near a rough boundary \cite{pawlak_maccready_2002}.

\section*{Conclusion}

In the present study, we have highlighted that the flow associated with a finite-amplitude water wave can induce separation of the bottom boundary layer in the vicinity of irregular regions of the topography. We have provided evidence that it also occurs for smooth boundaries with large, yet finite, curvature. Specifically, for a given Reynolds number $Re$ and a given wave amplitude, there exists a curvature threshold beyond which separation is initiated. As $Re\to +\infty \, ,$ this threshold tends to zero. The resulting perturbation persists in the vanishing-viscosity limit, thereby preventing convergence to the inviscid solution of the free-surface irrotational Euler equations. The singular nature of this limit for water waves constrains the range of validity of irrotational models. In particular, the vorticity shed from the bottom boundary layer significantly influences the shape of the free surface. This effect is especially pertinent in nearshore ocean dynamics, where vortices can rapidly propagate from the bed to the surface.

Long-wave models which accurately describe rapidly-varying bathymetries can also be derived from Euler's free-surface system without assuming irrotationality the flow \cite[\textit{e.g.}][]{clamond_dutykh_2011, dutykh_clamond_2016}. However, such models still fail at capturing the viscosity-based vorticity generation mechanism at hand.

In this study, we have focused on two-dimensional flows, which constitute a reasonable approximation for water waves in the nearshore regime. Naturally, the boundary layer separation described above is expected to extend to three-dimensional flows \cite{squire_1933}. In 3D, {counter-rotating vortices} correspond to pairs of vortex filaments with opposite vorticity. In this context, it is well established—from studies of wingtip vortices—that long-wavelength inviscid instabilities of these vortex lines can occur \cite{Crow70}, potentially leading to a cascade of complex vortex reconnections \cite{SciAdv20, PRF21}. However, such instabilities develop on a timescale that is long compared to the timescale associated with the propagation of the vortex pair. Therefore, the pronounced influence of vortex pairs on the free surface is expected to persist in three-dimensional flows.\\

\noindent\textbf{Acknowledgements.} We are grateful to Stephen Childress and Christophe Lacave for most useful discussions during the preparation of this work. We are grateful to the two anonymous referees for their insightful suggestions which helped improve the present work. We would also like to acknowledge the help of Georges Sadaka to implement the AMR procedure correctly in FreeFEM. All simulations have been carried out with help from the MCMeSU supercomputing centre.\\

\noindent\textbf{Data accessibility and reproducibility.} The FreeFEM scripts from which the numerical simulations visible in this work have been made is available at \href{https://gitlab.com/ariquier-research/water-waves-step}{\texttt{gitlab.com/ariquier-research/water-waves-step}}. All data generated using them can be found at \href{https://doi.org/10.57745/N7QK7T}{\texttt{doi:10.57745/N7QK7T}}.\\

\noindent\textbf{Competing interests.} The authors report no conflict of interests.\\

\printbibliography

@article{dutykh_clamond_2016,
    author = {Dutykh, D. and Clamond, D.},
    title = {Modiﬁed shallow water equations for signiﬁcantly varying seabeds},
    journal = {Applied Mathematical Modelling},
    year = {2016},
    volume = {40},
    pages = {9767–9787}
}

@article{viotti_dutykh_dias_2014,
    author = {Viotti, C. and Dutykh, D. and Dias, F.},
    title = {The conformal-mapping method for surface gravity waves in the presence of variable bathymetry and mean current},
    journal = {Procedia IUTAM},
    year = {2014},
    volume = {11},
    pages = {110-118}
}

@article{clamond_dutykh_2011,
    author = {Clamond, D. and Dutykh, D.},
    title = {Shallow water equations for large bathymetry variations},
    journal = {J. Phys. A: Math. Theor.},
    year = {2011},
    volume = {44},
    number = {332001}
}

@article{nguyen_etal_2011,
    author = {Nguyen, R. and Farge, M. and Schneider, K.},
    title = {Energy Dissipating Structures Produced by Walls in Two-Dimensional Flows at Vanishing Viscosity},
    journal = {Phys. Rev. Letters},
    year = {2011},
    volume = {106},
    number = {184502}
}

@article{pawlak_maccready_2002,
    author = {Pawlak, G. and MacCready, P.},
    title = {Oscillatory flow across an irregular boundary},
    journal = {J. Geophys. Res.},
    year = {2002},
    volume = {107},
    number = {C5}
}

@InProceedings{kato_1984,
    author="Kato, T.",
    title="Remarks on Zero Viscosity Limit for Nonstationary Navier-Stokes Flows with Boundary",
    booktitle="Seminar on Nonlinear Partial Differential Equations",
    year="1984",
}

@article{kelliher_2007,
    author          = {Kelliher, J. P.},
    journal         = {Ind. Uni. Math. J.},
    number          = {4},
    title           = {On Kato’s conditions for vanishing viscosity},
    volume          = {56},
    year            = {2007},
    pages           = {1711-1721}
}

@article{Crow70,
    author = {Crow, S. C.},
    title = {Stability theory for a pair of trailing vortices},
    journal = {AIAA Journal},
    volume = {8},
    number = {12},
    pages = {2172--2179},
    year = {1970}
}

@article{lin_huang_2010,
    author = {Lin, M.-Y. and Huang, L.-H.},
    title = {Vortex shedding from a submerged rectangular obstacle attacked by a solitary wave},
    journal = {J. Fluid Mech.},
    year = {2010},
    volume = {651},
    pages = {503-518}
}

@article{Hasselmann68,
    author = {Hasselmann, K. and Collins, J. I.},
    year ={1968},
    title={Spectral dissipation of finite-depth gravity waves due to turbulent bottom},
    journal={J. Mar. Res.},
    volume={26},
    pages={1-12}
}

@article{PRF21,
  title = {Cascades and reconnection in interacting vortex filaments},
  author = {Ostilla-M\'onico, R. and McKeown, R. and Brenner, M.P. and Rubinstein, S.M. and Pumir, A.},
  journal = {Phys. Rev. Fluids},
  volume = {6},
  issue = {7},
  pages = {074701},
  numpages = {21},
  year = {2021},
  month = {Jul},
  publisher = {American Physical Society},
}

@article{Lowe07,
author = {Lowe, R.J. and Falter, J.L. and Koseff, J.R. and Monismith, S.G. and Atkinson, M.J.},
title = {Spectral wave flow attenuation within submerged canopies: Implications for wave energy dissipation},
journal = {J. Geophys. Res.: Oceans},
volume = {112},
number = {C5},
year = {2007}
}

@book{feynman_1965,
  author = {Feynman, R. P. and Leighton, R. B. and Sands, M.},
  publisher = {Addison-Wesley},
  title = {{The Feynman Lectures on Physics; Vol. I}},
  year = {1965},
}

@article{squire_1933,
    author = {Squire, H. B.},
    title = {On the Stability for Three-Dimensional Disturbances of Viscous Fluid Flow between Parallel Walls},
    journal = {Proc. R. Soc. Lond. A},
    year = {1933},
    volume = {142},
    number = {847},
    pages = {621-628}
}

@article{lagerstrom_1975,
    author = {Lagerstrom, P. A.},
    title = {Solutions of the {N}avier-{S}tokes Equation at Large {R}eynolds Number},
    journal = {SIAM J. Appl. Math.},
    year = {1975},
    volume = {28},
    number = {1},
    pages = {202-214}
}

@book{sychev_etal_1998,
    author = {Sychev, V. V. and Ruban, A. I. and Sychev, V. V. and Korolev, G. L.},
    title = {Asymptotic Theory of Separated Flows},
    publisher = {Cambridge University Press},
    year = {1998}
}

@article{SciAdv20,
    author = {McKeown, R. and Ostilla-Mónico, R. and Pumir, A. and Brenner, M.P. and Rubinstein, S.M.},
    title = {Turbulence generation through an iterative cascade of the elliptical instability},
    journal = {Science Advances},
    volume = {6},
    number = {9},
    pages = {eaaz2717},
    year = {2020},
}

@book{lamb_1932,
  author         = {Lamb, H.},
  publisher      = {Cambridge University Press},
  title          = {Hydrodynamics},
  year           = {1932}
}

@article{grue_1992,
    author = {Grue, J.},
    title = {Nonlinear water waves at a submerged obstacle or bottom topography},
    journal = {J. Fluid Mech.},
    year = {1992},
    volume = {244},
    pages = {455-476}
}

@article{grue_2017,
    author = {Grue, J. and Kolaas, J.},
    title = {Experimental particle paths and drift velocity in steep waves at finite water depth},
    journal = {J. Fluid Mech.},
    year = {2017},
    volume = {810},
    pages = {R1}
}

@article{friedrichs_1944_mollifiers,
    author = {Friedrichs, {K.O.}},
    title = {The identity of weak and strong extensions of differential operators},
    journal = {Trans. Amer. Math. Soc.},
    volume = {55},
    year = {1944},
    pages = {132-151}
}

@book{Batchelor_1967, 
    title={An Introduction to Fluid Dynamics}, 
    publisher={Cambridge University Press}, 
    author={Batchelor, G.K.}, 
    year={1967}
}

@article{CraigSulem1993,
  title={Numerical simulation of gravity waves},
  author={W. Craig and C. Sulem},
  journal={J. Comp. Phys.},
  year={1993},
  volume={108},
  pages={73-83},
}

@article{Johnson_1973, 
    title={On the development of a solitary wave moving over an uneven bottom}, 
    volume={73}, 
    number={1}, 
    journal={Math. Proc. Camb. Phil. Soc.}, 
    author={Johnson, R. S.}, 
    year={1973}, 
    pages={183–203}
}

@article{KdV1895,
    author = {{D. J.} Korteweg and G. de Vries},
    title = {{XLI.} On the change of form of long waves advancing in a rectangular canal, and on a new type of long stationary waves },
    journal = {The London, Edin., and Dublin Phil. Mag. and J. of Sci.},
    volume = {39},
    number = {240},
    pages = {422-443},
    year = {1895},
    publisher = {Taylor \& Francis},
}

@article{Green_Naghdi_1976, 
    title={A derivation of equations for wave propagation in water of variable depth}, 
    volume={78},
    number={2}, 
    journal={Journal of Fluid Mechanics}, 
    author={Green, A. E. and Naghdi, P. M.}, 
    year={1976}, 
    pages={237–246}
}

@article{SaintVenant1871,
    author = {{Barré de Saint-Venant}, {A. J. C.}},
    title = {Théorie du mouvement non permanent des eaux, avec application aux crues des rivières et à l'introduction de marées dans leurs lits},
    journal = {Comptes Rendus Acad. Sci.},
    volume = {73},
    pages = {147-154 and 237-240},
    year = {1871}
}

@article{Riquier_Dormy_2024,
    author = {Riquier, A. and Dormy, E.},
    journal = {J. Fluid Mech.},
    title = {Numerical study of a viscous breaking water wave and the limit of vanishing viscosity},
    volume = {984},
    year = {2024},
    pages = {R5},
}

@book{johnson,
	author = {R. S. Johnson},
	title = {A Modern Introduction to the Mathematical Theory of Water Waves},
	year = {1997},
	publisher = {Cambridge University Press},
}

@article{vonKarman1911,
    author = {von Kármán, T.},
    journal = {Nachrichten von der Gesellschaft der Wissenschaften zu Göttingen, Mathematisch-Physikalische Klasse},
    pages = {509-517},
    title = {{\"U}ber den Mechanismus des Widerstandes, den ein bewegter Körper in einer Flüssigkeit erfährt},
    year = {1911},
}

@article{FreeFem,
	AUTHOR = {Hecht, F.},
	TITLE = {New development in {FreeFem++}},
	JOURNAL = {J. Numer. Math.},
	FJOURNAL = {Journal of Numerical Mathematics},
	VOLUME = {20}, YEAR = {2012},
	NUMBER = {3-4}, PAGES = {251--265},
	ISSN = {1570-2820},
	MRCLASS = {65Y15},
	MRNUMBER = {3043640},
	URL = {https://freefem.org/}
}

@book{lannes2013,
	TITLE = {{The Water Waves Problem: Mathematical Analysis and Asymptotics}},
	AUTHOR = {Lannes, D.},
	publisher = {American Mathematical Society},
	SERIES = {Mathematical Surveys and Monographs},
	VOLUME = {188},
	YEAR = {2013},
}

@article{ALE_HIRT1974,
    title = {An arbitrary Lagrangian-Eulerian computing method for all flow speeds},
    journal = {Journal of Computational Physics},
    volume = {14},
    number = {3},
    pages = {227-253},
    year = {1974},
    issn = {0021-9991},
    author = {C. W. Hirt and A. A. Amsden and J. L. Cook}
}

@article{LonguetHiggins1953,
 author = {M. S. Longuet-Higgins},
 journal = {Phil. Trans. Roy. Soc. London A},
 number = {903},
 pages = {535-581},
 title = {Mass Transport in Water Waves},
 urldate = {2023-05-19},
 volume = {245},
 year = {1953},
note = {see page 537}
}

@article{DormyLacave2024,
    title={Inviscid Water-Waves and interface modeling}, 
    author={E. Dormy and C. Lacave},
    year={2024},
    journal={Quar. Appl. Math.},
    volume = {82},
    pages = {583-637},
}

@article{Raval2009,
    author = {Raval, A. and Wen, X. and Smith, {M. H.}},
    title = {Numerical simulation of viscous, nonlinear and progressive water waves},
    journal = {J. Fluid Mech.},
    year = {2009},
    volume = {637},
    pages = {443-473}
}

@article{Zakharov1968,
    author = {{V. E.} Zakharov},
    title = {Stability of periodic waves of finite amplitude on the surface of a deep fluid},
    journal = {Journal of Applied Mechanics and Technical Physics},
    year = 1968,
    volume = 9,
    pages = {190-194}
}

@article{Deike2015,
    author = {Deike, L. and Popinet, S. and Melville, {W. K.}},
    title = {Capillary effects on wave breaking},
    journal = {J. Fluid Mech.},
    year = {2015},
    volume = {769},
    pages = {541-569}
}

@article{Masmoudi_Rousset_2017,
    author = {Masmoudi, N. and Rousset, F.},
    title = {Uniform Regularity and Vanishing Viscosity Limit for the Free Surface Navier–Stokes Equations},
    journal = {Arch. Rational Mech. Anal.},
    year = {2017},
    volume = {223},
    pages = {301–417}
}

\appendix

\section{Numerical parameters and convergence}
\label{sec:appendix}

In this appendix, we address technicalities regarding the thinness of the unstructured grids used to obtain the results presented in this work. We also discuss the numerical accuracy in resolving the bed's boundary layer structure and the separation.

\begin{figure}[h]
    \centering
    \includegraphics[width=\textwidth]{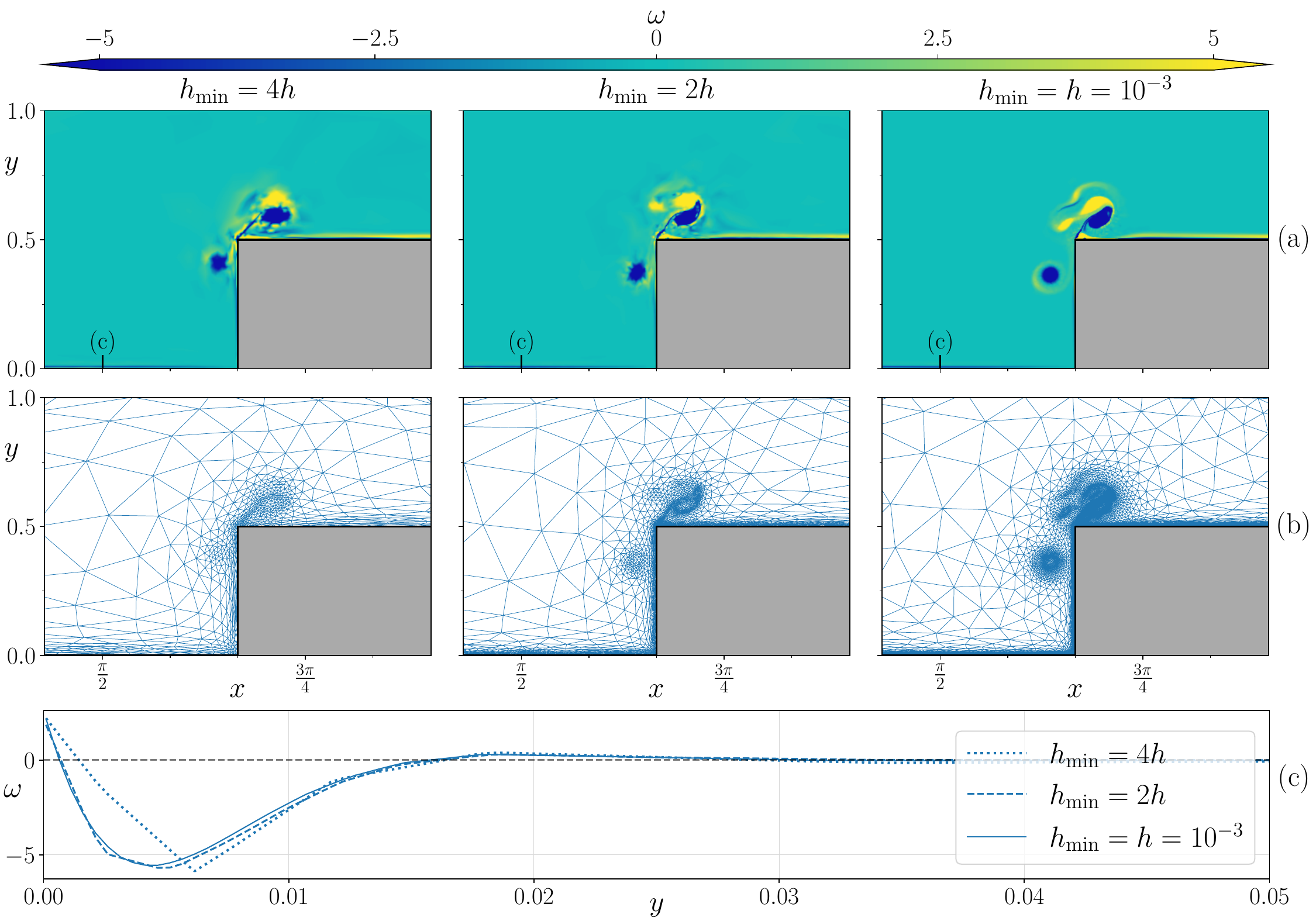}
    \caption{Numerical convergence of the $Re = 10^5$ simulation with a sharp step at fixed time $t = 10$. The number of elements used to discretise the boundary has been halved two times, along with the AMR precision (see table \ref{tab:numerical_params}), yielding the coarser (middle column) and coarsest (left column) results. (a) Resulting vorticity in the vicinity of the left corner of the salient edge. (b) Visualisation of the corresponding mesh. (c) Value of the vorticity in the boundary layer at $x = \frac{\pi}{2}$ (black lines in panels (a)).}
    \label{fig:numerical_convergence}
\end{figure}

A summary of the numerical parameters used to build the initial meshes is available in table \ref{tab:numerical_params}. As FreeFEM's native Adaptative Mesh Refinement (AMR) procedure \cite{FreeFem} gets called, the number of elements used to discretise the bottom boundary is changed to resolve the vortical structure. The metric used by the AMR algorithm is created taking into account the hessian of the fluid's velocity, thus ensuring that regions of strong vorticity gradient are resolved using small triangles, along with the distance to the top and bottom boundaries. In this way, we can make sure that the bed's boundary layer always consists in at least 10 triangles in the direction perpendicular to the boundary (see Fig. \ref{fig:numerical_convergence}c).

\begin{table}[!h]
    \centering
    \caption{Numerical parameters used in the simulations carried out in the present study, provided for replicability. $N_\mathrm{top}$, $N_\mathrm{bot}$ and $N_\mathrm{side}$ correspond, respectively, to the number of elements initially used to discretise the free surface, the bathymetry and the two periodic vertical boundaries. The parameter $h_{\mathrm{min}}$ is the minimum triangle diameter used by the AMR algorithm. Along the simulations, the AMR changes the number of elements used to discretise the bottom boundary, effectively changing the value of $N_\mathrm{bot}$. It leaves, however, both $N_\mathrm{top}$ and $N_\mathrm{side}$ unchanged. In all cases, the initial condition has been calculated using the values $L = 2\pi$, $A = 0.1$, $k = 1$, $h_b = 0.5$ and $\ell_b = L/3$ (before mollification when relevant).}
    \label{tab:numerical_params}
    \begin{tblr}{X[l,m]Q[c,m]Q[c,m]|Q[c,m]Q[c,m]Q[c,m]Q[r,m]}
    \hline[1pt]
        \textbf{Description} & $Re$ & $\varepsilon$ & $N_\mathrm{top}$ & $N_\mathrm{bot}$ & $N_\mathrm{side}$ & $h_{\mathrm{min}}$ \\
    \hline[1pt]
         & $10^2$ to $10^4$ & -- & $2000$ & $900$ & $250$ & $2\cdot 10^{-3}$ \\
        Sharp step (sec. \ref{sec:rect_step}) & $10^{4.5}$ & -- & 2500 & 900 & 250 & $1.5\cdot 10^{-3}$ \\
         & $10^5$ & -- & 2500 & 1350 & 250 & $1\cdot 10^{-3}$ \\
    \hline[dotted]
        Mollified step (sec. \ref{sec:smoothness}) & $10^5$ & $0.1$ to $1.0$ & 2500 & 1350 & 250 & $1.5\cdot 10^{-3}$ \\
    \hline[dotted]
        Num. conv. $h_{\mathrm{min}} = 2h$ & $10^5$ & -- & 1250 & 675 & 125 & $2\cdot 10^{-3}$ \\
        Num. conv. $h_{\mathrm{min}} = 4h$ & $10^5$ & -- & 625 & 340 & 63 & $4\cdot 10^{-3}$ \\
    \hline[1pt]
    \end{tblr}
\end{table}

To estimate whether the numerical solutions showed in this work remain close to the solutions of system (\ref{eq:Navier-Stokes}--\ref{eq:stress-free}), we provide evidences that the boundary layer separation depicted in this study would not differ should we decrease the mesh's size any further. This is the purpose of Fig. \ref{fig:numerical_convergence} in which the numerical simulation with a Reynolds number $Re = 10^5$, used to obtain the results shown in Figs. \ref{fig:compare_interfaces}--\ref{fig:ensrophy}, is compared with the ones obtained increasing twofold and fourfold the size of the elements along with the acceptable error of the AMR routine. The top panels $(\mathrm{a})$ show the vorticity computed for these simulations, which converges (slowly) to a grid-independent vortical structure. The panel $(\mathrm{c})$ shows the boundary layer vorticity at $x=\frac{\pi}{2}$. One sees that the optimal grid resolution to fully resolve the boundary layer is already attained in the $h_\mathrm{min} = 2h$ simulation. However decreasing further the mesh's size is necessary to capture the separated vorticity accurately (Fig. \ref{fig:numerical_convergence}a).

\end{document}